# Health Matters:
# Human Organ Donations, Sales, and the Black Market

Michael Hentrich

**ABSTRACT.**

*In this paper I explore the human organ procurement system. Which is better for saving lives and limiting black market use, the present altruistic system of donations or a free and open sales market? I explain that there is a risk with maintaining the present system, the altruistic vision, and that people may die who might otherwise live if the sale of organs was permitted. But there is no guarantee that permitting organ sales would effectively address the current supply-side shortage and global use of the black market. In addition to discussing the implications of these procurement systems, I look at methods to increase organ donations and I explore the differences between presumed and explicit consent. Ultimately, I conclude that the altruistic donation system, bolstered by the addition of a policy of presumed consent and appropriate financial incentives, is a better choice than a legal sales market in spite of its shortcomings.*

**METHODOLOGY**

The human organ procurement system is a much debated and controversial topic. With reference to the sociological and economic dimensions of existing organ networks and procurement policies, this paper aims to (1) explain the allure and logic of altruism as opposed to a free sales market, (2) understand the problems associated with the black market, and (3) make a policy judgment. By exploring the issue in depth I hope to provide a framework with which to view the issue neutrally. I address whether it ought to be illegal to sell organs, as it is for most of the world, what that means for donors and recipients, and whether the sale of organs is, or could be, safe and efficient.

With better understanding of the nature of organ rejection and development of the techniques and technology necessary to perform transplantation, the first few successful organ transplants with live human subjects took place throughout the 1960s. Since then, the human organ procurement system has been the source of both physical and emotional trauma for a great many people, especially candidates for transplant procedures. The nearly unanimously implemented system that exists at present is a simple voluntary donation mechanism, where organs are given either after death or during life in the case of those that are not needed by the donor to survive (one of two kidneys and portions of the lungs and liver). Donating to a specific person sidesteps the issue of waiting on a long list for an organ and is more frequently practiced between family members, whereas donating to a non-relative is not as commonplace. Waiting for an organ from an unrelated donor can take months or even years depending on the organ and availability. Thousands of people die every year waiting for a kidney, heart, liver, or other organ, and there are a growing number of people on the waiting list for transplants.

An undisputed characteristic of the existing organ procurement system is that the demand is much greater than the supply. As a result, the price of black market organs is driven up[1], and law-abiding citizens on the waiting list are often not helped. While many of the organs donated to unrelated people are from individuals who decided while they were alive to donate their organs upon death, in spite of their noble efforts there is still a supply-side shortage (Banks 1995).

According to the U.S. Department of Health & Human Services, in 2011 in the United States alone 112,708 people were waiting for an organ[2], and someone was added to the waiting list every eleven minutes. This is up from 68,000 Americans waiting for an organ in the prior decade, when an average of twelve people on these waitlists died every day (Harris and Alcorn 213, 2000-2001). Compare this to the average of 75 people who receive an organ transplant every day in the United States, yielding a national recipient total of 27,375 people per year and leaving a shortage of 85,333 organs per year[3]. While demand is still

1 The effect of supply-side shortage on price is particularly pronounced in the black market. However, supply shortage also increases the expenses that are sometimes reimbursed in countries which permit financial compensation, such as travel expenses, lost wages, hospitalization, and extended health care (Harris and Alcorn 2000-2001, Becker and Elias 2007).

2 It is important to note that such a figure may be inflated. The U.S. Department of Health & Human Services notes that "one of the most confusing statistics is the number of persons waiting for a transplant. Patients are allowed to register at multiple transplant centers so you may see a higher number if you count 'registrations' rather than 'candidates.'"

3 In terms of the racial breakdown of donors, white donors accounted for 67 percent of all donations in 2008, while amounting to 63.7 percent of the population in the United States. Compare that to black donors accounting for 16 percent of all donations in the same year, and amounting to 12.3 percent of the population. Hispanics account for 8.7 percent of the population and 14 percent of all donations, and Asians account for 4.8 percent of the population and 2.5 percent of all donations. At the same time, the national waiting list is disproportionately made up of whites: 45 percent are whites, 29 percent blacks, 18 percent Hispanics, and 6 percent Asians (U.S. Department of Health & Human Services).

increasing, supply has remained steady in recent years. Observe that while organ donations increased in absolute numbers from the late 1980s up until the mid-2000s, donation rates did not increase or decrease significantly from 2004-2010 (shown in Figure 2, below). While deceased donor rates have doubled since 1988, living donor rates have more than tripled. Of course, part of the increase in donor rates is an active response to the increased waiting list. Yet donation rates have increased alongside, but not in proportion to, increased population. At the same time, 30.7 percent of kidney donor recipients have died within five years of receiving their transplant, as well as 25.1 percent of heart recipients, 26.2 percent of liver recipients, and 45.6 percent of lung recipients (U.S. Department of Health & Human Services).

The black market for organs refers to the criminal act of offering organs for sale when sale is illegal; such a market exists partially if not primarily as a response to the global supply-side shortage. Hence the black market is a problem in developed and developing nations alike. Problematically, poor citizens within developing countries are often the ones selling organs, especially kidneys, through the black market; even children have sold their organs.[4] Furthermore the potential long term health risks are seldom fully explained to, or appreciated by, the organ seller. According to one report, individuals are compensated anywhere from $6,000 to $10,000 plus airfare on the high end ($800 on the low end) for one of their kidneys, which is then sold by a black market middleman for anywhere up to

4 In one documented case, a seventeen-year-old boy in China told a local television station that he sold his kidney for the money to buy an iPad (Patience 2011).

$100,000 (Corwin 2011).[5] In the United States, a black market has existed for tissue, where black market brokers made deals with funeral home owners and directors to harvest tissues and parts from bodies in their possession without the consent of the individual or family. These were then sold for a profit to researchers and doctors; false reports were sometimes created to cover up the true cause of death and the fact that the tissues could be diseased.

The public stigma that surrounds the black market is enough to convince many to wait it out legally, even if it means death (Goodwin 2006). In certain cases, potential organ recipients are also deterred by the uncertainties surrounding a black market transplant in a foreign country, where the quality of both the organ and medical care is questionable. At the same time, the presence of the black market actively undermines the legal organ donation and procurement system, especially because black market goods have been acquired illegally and sometimes without consent. For example, black market organs may have been taken from a patient while undergoing other surgical procedures. This happens with kidneys and parts of lungs since the patient can survive without them, and because the illegal organ theft goes initially undetected (Goodwin 2006). It has even been reported, but with little hard evidence, that people have been killed for their organs by black marketers. Such illegal commerce exists and exploits the

5 According to another source, an estimated 800 kidneys were being sold every year in the Philippines and transplanted to foreigners before a ban went into effect in 2008, with people in developed countries travelling to poorer countries to receive these organs for a premium. Poor Filipinos selling a kidney received as little as $2000, whereas the hospitals performing the transplants were involved in a lucrative business, generating $50,000 to $80,000 per transplant (Abou-Alsamh 2009). Brokers in Yemen reportedly received as much as $60,000 for kidneys procured from poor Yeminis and Egyptians, who typically received as little as $5000 and who were often robbed of this money on their way back home.

poor, robbing them of vital bodily resources. Nonetheless, while the current system of altruism allows for the global presence of a black market for organs, there is no clear evidence that a sales market would shut down such operations.

What are the arguments for a non-commoditized altruistic system which depends on donations and not on compensation? Relying on altruistic motives is indeed compelling. For many cultures, the body is regarded as a sacred entity and donating an organ honors this sacralization of bodily resources, whereas the sale of organs is regarded by most cultures to be taboo. And so one argument for maintaining the status quo is that, out of the range of policies that could be implemented, a ban on organ selling and a procurement system based on donations only is the most culturally acceptable and therefore the most politically viable policy.

The gift-relationship can be summed up as both an explicit rationale by which donation-making decisions are made and the supply and demand side explanations that lead to these decisions. Richard Titmuss in *The Gift Relationship* (1971) writes on the role of altruism for meeting the demand for blood in the United States and Great Britain. The right to give implies that a choice must be made on behalf of the recipient. The present system everywhere except in Iran and a few other countries encourages altruism, seeing it as a principal and acceptable motivation and incentive. For family members, it is undoubtedly the case that donations are motivated by the fear of losing their loved ones, though the same motivation does not apply to donations for strangers. But even between strangers, trading organs for money is unsettling because it implies that a price can be placed on a human good. Additionally, the marketization of organs and blood could have the undesired effect of crowding out the supply of former donors (now paid suppliers), because the act of donating an organ would become less morally significant[6] (Titmuss 1971).

In defense of the status quo, Kieran Healey in *Last Best Gifts* (2006) offers a brilliant account of the role played by various institutions and actors who are involved in organ procurement and in shaping the gift narrative. He notes that procurement organizations presently promote altruism and the donation of organs, and agrees with some experts that monetizing the organ market would produce a risky environment on the supply side, worsening conditions that are already bad. Put another way, the concern, following Titmuss' logic, is that monetization destroys altruistic motives and thus decreases supply. Healey writes about commoditization (or monetization) as follows:

> We get what we wish for. If we talk of blood as if it were a commodity, then people will come to commodify it in practice. By instituting a market for blood or organs, people orient themselves toward these goods in a new way. The rational calculus of costs and benefits comes to override alternative ways of thinking about the value of what is being exchanged (Healey 11, 2006).

If an organ sales market was adopted, Healey argues that the key to its success is fairness of the exchange, such that the poor are not exploited and the many that

6 Titmuss explains, as evidence of the crowding out effect, that the amount of blood donated and the number of people who donated increased in the United Kingdom, where the sale of blood was prohibited, while these numbers declined in the United States, where the sale of blood was allowed. And so we observe, the sale of blood wears away the incentive to give (Titmuss 1971).

need organs are able to afford and receive them (Healey 2006).

What are the various arguments and critiques against the altruistic procurement system and what is recommended to stand in its place? Much of the recent literature on the gift-relationship contends that some alteration of the status quo is to be preferred. The major problem associated with the status quo, according to legal scholar Gloria Banks (2011), is that the legal and donor-based organ procurement system suffers from scarcity and a supply-side shortage. On that note, does the gift- relationship limit or facilitate illegal black market behavior and would it be a better policy to commercialize the sale of organs or regulate incentives?

Ben-David (2005) explains further that the inadequacies of the altruistic vision include its inability to provide proper incentives for organ donation. Decisions to donate have not kept up with medical advancements via technology. On a separate but related point, he argues:

> The transplantation of organs, which at first sight appears to be just a technical medical action, is first and foremost a socio-cultural act, in which two aspects of exchange operate. One is concrete and conscious, while the other is abstract and subconscious (Ben-David 150, 2005).

The conscious act of exchange, or the literal transference of an organ from one person to another, is contrasted to the symbolic act of exchange and the ideological values of life and death that guide the decision-making process for society and for individuals. The taboo on selling organs is based primarily on these ideological values, which stand in the way of commoditization; the argument is made that selling organs would be a more cost-effective policy and would increase supplies.

The proposed alternative systems come in two forms – incentive-based governmental regulation and market-based sales. Writing about the supply shortage for organ donations and the effect of governmental regulation, Curtis Harris and Stephen Alcorn (2000-2001) discuss the economic incentives created by government action and inaction for various actors in the organ procurement system, including Organ Procurement Organizations (OPOs). The presence of OPOs has altered industry structure[7] in the United States and elsewhere. OPOs serve to manage the organ procurement process on a large scale and arrange for individual donations on the small scale; they are the gatekeepers of the organ procurement system and transfer process. OPOs, together with doctors and politicians, are the major actors in the sociology of the organ procurement system with influential views about, and vested interests in, the current system.

The role of technology is, in part, to contextualize the pragmatics and praxis of a system. Technology operates as a great resource to many, but is only applicable where basic human labor serves to make use of its essential properties. Transplant technology, put another way, serves to preserve organs and to aid in the transplant itself but only if there are enough doctors to perform the transplant and only if there are enough organs in supply. On the subject of technology,

7 The National Organ Transplant Act of 1984, amended in 1988 and 1990, allowed the Department of Health and Human Services in the United States to organize the formation of OPOs including the Organ Procurement and Transplantation Network and the Scientific Registry of Transplant Recipients. According to the Association of Organ Procurement Organizations, OPOs are "by federal law… the only organizations that can perform the life-saving mission of recovering organs from deceased donors for transplantation".

Barnett, Beard, and Kaserman (1993) critique the present system, and argue that doctors and medical professionals oppose organ sale technology because they have a financial stake in maintaining the status quo. Concerning the vested interests of such professionals, the authors write:

> The ongoing shortage of kidneys, hearts, livers, lungs, and other solid organs has significantly hampered the ability of physicians to bring improved life-saving transplant technology to patients suffering from a variety of debilitating and often fatal diseases… opposition to the formation of organ markets has been quite strong among transplant suppliers (both hospital and physician groups) (Barnett, Beard, and Kaserman, 1993).

One of the concerns associated with any organ procurement system is the many who are exploited by black market systems, where organs can be taken without full and fair consent or indeed any permission at all. In certain cases, especially when the donation is needed for a family member, the opposite may be true; Shanteau and Harris (1990) write:

> One possible explanation is that the earthly rewards for being a donor can only be experienced while living, not after death. With a relative, a donation would directly benefit a known other person. If a close relative's life is saved, then the donor would still be alive to share in the experience. A donation after death cannot produce a similar experience. Therefore, the opportunity to see the rewards of donation may provide an added motivation (Shanteau and Harris 65, 1990).

Contemporary sociologists including Michele Goodwin in *Black Markets: The Supply and Demand of Body Parts* (2006) have criticized the present system, pointing out that where there is a gift-relationship procurement system, a thriving black market also exists. Goodwin also argues that poor African American communities are especially ill-served by black markets. There are simply no guarantees that either regulation or marketization would alleviate black market use rates and supply-side shortages. Consider that a black market prospers even in Iran where the government regulates kidney pricing; black market use there (even for kidneys) has not been eliminated or even substantially limited.

I have referred to the sales of organs multiple times throughout this paper. With reference to such a system, Jason Altman (2007) discusses the possibility of an open international organ sales market permitting the marketization, free exchange, and regulation of organs among global nations. Does the creation of something like the European Union for organs have the unmet potential to yield greater numbers of organs available for transplant? Again, there are no guarantees as to the effects of implementing such a system. Altman is proposing a literal market, with payments and monetary incentives. Such a system could be detrimental to supply, however, because the appeal of altruism could be lost.

Schwindt and Vining (1986) add their voices to the call for a legal sales market similar to that suggested by Jason Altman. The authors argue that such a system would correct for supply-side shortages, writing as follows:

> The major roadblocks to increased transplantation are lack of donor organs and financing… For almost all organs, supply cannot keep up with demand, and demand is increasing. There is little hope that the current "altruistic" system will be able to keep up with either existing or emerging demand… market forces [are] a method of both generating an adequate supply of transplant organs and ensuring that the supply is efficiently used (Schwindt and Vining 483, 1986).

Rothman and Rothman (2006) further write that establishing a market for organs has already garnered significant support. They note that the problem with such a market has to do with how to avoid the ethical failures of the present system, and how to prevent the inefficiencies associated with crowding out, which would wear away at the moral fabric of a system in which the poor are exploited.

The market for blood and reproductive cells tells a quite different story than that of the organ procurement system; it is comparatively easy to donate these bodily resources, the donor does not risk his or her health by donating and these resources are replenished or at least not as necessary to the donor's health and survival as an organ. Feldschuh in *Safe Blood* (1998) explains one difference between blood transfusion and organ transplantation when he writes:

> If you have your own frozen stored blood available, proper full replacement can take place. A doctor who knows that you have stored your own risk-free blood should not hesitate to provide you with complete replacement when you lose blood (Feldschuh 160, 1998).

Giving blood is not the same as donating or selling an organ because the latter carries significantly greater risks to the donor (or seller) and because blood is a renewable resource, whereas organs (excluding the liver) are not. Blood is more readily available, sold more frequently, and sold legally in the United States; there is less of a taboo against doing so.

The prevailing practice in the United States and most other countries is to procure organs from donations, and to keep those in need of an organ waiting on a list for a legal, donated organ. The prevailing thought behind the arguments in favor of permitting organ sales or offering financial and other incentives for organ donations is that more organs will then be made available for transplant (Becker and Elias 2007). Of course, it should be noted that selling organs or adding incentives has the potential to increase the medical and administrative costs associated with the transplantation itself and the extensive care required following transplantation, because we live in a health care system where costs for almost everything related to health have greatly increased over time.

The implications of permitting the sale of organs also differs by country based on levels of wealth and cultural norms. The same policy decisions made in the United States and Kenya would have vastly different results. Global policy decisions about organ transplant made purely on a homogenous economic analysis could well be misguided by failing to account for cultural norms and differing social conditions (Kaserman 2002). In developing countries the formal institutions involved with organ transplant are also less advanced. There are fewer doctors in the related areas and fewer transplant organizations through which to organize a legal market. These conditions combine to leave developing countries open to poorly regulated markets, abuse of donors and sellers, and the existence of a black market for organs obtained in ways that may not be fair and legal (Goodwin 2006).

To date, organs come from cadavers and from living donors. The role of donor campaigns is significant, and those who lobby for organ donations save and have saved lives. I have alluded to the problem of crowding out in this paper; it simply means that one of the widely noted concerns with legalizing the sale of organs is that it would actually reduce the number of donors because the act of donating an organ would become less morally significant[8]. Organs would become just another commodity to be traded on the market, and the moral and ethical aspects of voluntary donation would be diminished in the eyes of those who would give their organs. The energy behind the voluntary donation campaigns might also be undermined, further reducing the volunteer or gift-related supply. The focus on economic factors must also, for these reasons, be evaluated in the broader social and cultural context (Barnett, Beard, Kaserman 1993, 669-678).

Iran is the only country that has adopted a formal, government-regulated system for kidney sales. One Iranian surgeon familiar with the program describes it as follows:

> There is no commercialism. There are no middlemen or companies to sell the kidneys. No patient can go and buy a kidney. There is no benefit to the transplant team. The operation is just part of the overall medical program at the university. There are no foreign recipients. Nobody can come to Iran to buy a kidney… There are no foreign donors… Rich and poor are transplanted equally. There is no discrimination. The donor is free to refuse the government reward (Transplant News 2003).

Rather than being sold in a market, kidney prices in Iran are highly regulated by the government (BNET 2003; Tober 2007, 151-70). The government pays the organ supplier, who also receives some money from either the recipient or a charity[9]. Since adopting this system, donations of organs in Iran have reportedly not declined and there are supposedly no waiting lists (for kidneys). However, others have questioned some of the data from Iran and some negative consequences to kidney sellers have been noted, including a shortened lifespan. Also there are questions as to whether the social context of Iran is critically different from that of Western countries, such that a similar program would not work as well in the United States and elsewhere.

---

8 The crowding out effect is discussed by Rothman and Rothman (2006) in *The Hidden Cost of Organ Sale*, when they write: "Since the 1970s, a group of economists and social psychologists have been analyzing the tensions between 'extrinsic incentives' – financial compensation and monetary rewards, and 'intrinsic incentives' – the moral commitment to do one's duty. They hypothesize that extrinsic incentives can 'crowd out' intrinsic incentives, that the introduction of cash payments will weaken moral obligations…It does suggest that a market in organs might reduce altruistic donation and overall supply." (Rothman and Rothman 1525, 2006).

Another way to increase donations and organ supplies without financial or other incentives is through presumed consent. In some European countries (including Austria, Belgium, France, Hungary, Poland, Portugal, and Sweden), when someone dies that person is presumed to have consented to donate any usable organs. This form of consent is overridden only if the person specifically makes it clear that he or she does not want to participate in donating organs after death. Such a system is referred to as an "opt-out" consent system. In these countries we know that organ availability increases alongside substantially increased donation rates (Ariely 2011). In the United States, we have the opposite system – a person has to "opt-in" to organ donation after his death and make that consent known prior to his death.

Individuals in countries like the U.S. which practice opt-in consent are asked by their Departments of Motor Vehicles to "Check the box below if you

9 The present system in Iran is regularized and regulated with set governmental restraints. It is not marketized because the price is not set by market demand.

*want* to participate in the organ donor program", whereas individuals in the countries which practice opt-out consent are asked to "Check the box below if you *don't want* to participate in the organ donor program". Yet this seemingly minor difference in form yields hugely significant results (Ariely 2011, Zink 2005).

Just over one-third of all organs taken after death are able to be transferred into another body. The benefit of presumed consent, when doctors take all the necessary precautions not to prematurely declare somebody as dead, is that it avoids the moral quandary associated with the sale of organs. If presumed consent programs were adopted and implemented in all of the countries where they do not exist at present, this would lead to a significant increase in the number of organs available for transplant and prolong the lives of a great number of individuals (Ariely 2011, Zink 2005).

Religious practices and institutions also play a part in the donation process. Religious approval can be crucial to those potential donors who rely on their religion to tell them that donations are allowed and morally appropriate. In that respect Kieran Healey notes:

> But insofar as the donor is dead and there is no money involved, most Christian church authorities are not against organ donation… Orthodox Judaism, however, has had more trouble assimilating organ donation to existing law and practice. There is more opposition to organ donation, and religious authority is more divided on the issue…Once church leaders had satisfied themselves that brain death was a valid concept, they no longer

> opposed organ donation. All major Christian denominations came to support donation and transplantation as a morally valuable activity. (Healey 31-32, 2006.)

The fact that there would probably be serious religious backlash to the marketization of organs must be included in the calculation of whether such a market should be permitted. Such a change in the way organs are procured would reflect a fundamental change in the morality of the act in the eyes of many religious authorities.

In addition to making use of presumed consent and being sensitive to cultural and religious attitudes, donations could be increased by making payments to donors in ways that take into account the expressive role of money. Sociologists studying this issue have determined that there is invariably a place for paying donors so long as you do it the right way. For example, transplant institutions (including governments that want to encourage donations) might be able to offer certain forms of financial incentives to the families of dead persons in order to promote donations, so long as these incentives are sensitive to the expressive role of money. One suggested solution is to offer money for funeral expenses to families who agree to organ donations, enabling families to honor their deceased relative. Such a solution allows for honorific exchange as opposed to market exchange, and further incentivizes those who supply an organ (Becker and Elias 2007).

Eurotransplant, an independent and collaborative organization that supports international donor exchange, has suggested as an incentive that living donors be placed higher on the transplant waiting list if they ever need a kidney later on in their lives. In Israel (as in Iran), there is a system where donors receive a fixed amount of compensation from the government. There is also a government-sponsored education budget to inform the public about organ donation[10]. In tune with the religious spirit of the nation and to protect the sacralization of organs, the laws in Israel effectively ban the trade of organs while also facilitating the aforementioned legal alternative (Ratzlav-Katz 2011). Additional incentives proposed in Israel include government-granted income for donors, municipal tax exemptions, free public transportation, tuition grants, free passes to public parks, and the granting of an honorific certificate.

Within the framework of a non-market system, another idea for increasing the supply of organs and further incentivizing donations is through a futures market (Healey 2006). The idea is that people agree while they are still alive to donate their organs when they die, in exchange for a small payment they receive while they are still alive. An alternative futures market system is one where individuals agree to donate their organs after death in exchange for payment made at the time of their death to their relatives or to a charity. These compensation systems avoid the many complications and potential abuses of taking organs from a live donor (Healy 2006, 35-36).

Financial compensation for donations could also be increased by broadening the kind of expenses that are reimbursed to donors. An organ donor should be guaranteed reimbursement for all medical expenses associated with donation and transplant. But in addition to medical expenses, a living donor could be also be reimbursed for any travel or other incidental expenses involved and for wages he would have earned while preparing for and recovering from surgery. A donor could also be guaranteed reimbursement for any future health care needs or future income lost because of the donation (Becker and Elias 2007).

---

10 The public's unwillingness to donate organs in Israel and other countries is related to the fear that doctors will declare a patient's death prematurely in order to collect organs and even profit from them. For that reason polls in Israel show that less than 50 percent of the population is willing to donate. Money spent on public education has the potential to increase that number.

The trouble with adding compensation and incentives to organ donation is with maintaining the status of the transaction as a gift rather than as a sale. At some point, do we not cross a threshold where the financial incentives or reimbursed expenses amount to organ sale? Put another way, while the gift relationship with smaller payments protects the sacralization of organs, if the financial compensation is too high people may simply consider the transaction to be a sale (Goodwin 2006). What is at stake is the status of altruism associated with the gift relationship, because organ donation has come to be viewed as an altruistic, sacred activity (Mocan and Tekin 2007, 2527-538; Tilney 2003-2004).

The shortage of organs and the life and death consequences associated with getting an organ create the circumstances that give rise to a black market. The problems involved with a black market include abuse and exploitation of poor donors, quality issues, and the perpetuation of a system where the wealthy receive a disproportionate share of available organs, both by utilizing the black market and devoting more resources to finding a donated organ. Black markets have a larger presence in peripheral and semi-peripheral countries, where there is a ready supply of individuals who are sufficiently desperate for money that they are prepared to risk their health by selling organs (Goodwin 2006).

Even if organ sales were legalized, it is not clear what effect this would actually have on the shortage of organs for transplant. Predictions of price and market behavior are highly uncertain. If sales of kidneys were permitted, would the supply of kidneys catch up with the demand? What would happen to kidney prices? Unregulated, it is difficult to imagine that prices would be affordable to all those in need (Corwin 2011). Contemporary sociologists have been quick to suggest that the supply-side shortage may be cured by legalizing sales; the truth of such claims rests on whether the supply would increase enough to provide an organ for everyone in need (Corwin 2011, Ben-David 2005). Of course, demand can only increase inasmuch as there are people in need of organs, while the numbers who can give organs is practically limitless.

Contingency theorists contend that there are limits to organizing society because there is no single best way to make decisions; organizations need to remain flexible to respond to changing environmental demands. The appropriate form of organization depends on the kind of task or environment one is dealing with, and management must be free to adapt the organization to fit current circumstances. In the case of organ transplants, management consists of government officials and politicians who control the laws pertaining to organ procurement as well as OPOs and the doctors and hospitals who perform transplantation.

Recent advancements in medical technology used to preserve and transport organs demands precise and intentionally exacted management about the issue at hand; achieving greater awareness of an issue that has global significance is critical to changing the procurement system (Becker and Elias 2007, Harris and Alcorn 2000-2001). Another related point of contingency theory explicitly stresses the role of management because in any group it is the managers who make the decisions that most profoundly influence the rest of the population and who have the greatest control over circumstance.

Technology also has the ability to transform large-scale operations and efficiencies of any system. Revolutions of advancement have consistently worked wonders by transforming the labor force into more specialized groups, and the information revolution is no exception. Technology already has transformed parts of the organ transplant industry (specifically preservation and transportation)[11]. In the near future we may need fewer people to perform a transplant, especially if the transplant becomes mechanized and machines are able to help doctors perform procedures. Such advancements may increase efficiency, but to no great avail unless the availability of the organ is simultaneously increased. Technology does, however, remove one of the barriers associated with increasing supplies, which is to say that as transplants become technically easier and safer, there are fewer reservations about choosing to give.

11 It is now more efficient and effective to process organs, which saves time and resources and improves success rates by protecting the organ from deterioration.

While organized altruism theoretically has no limits, it produces a substantial shortage in practice (Becker and Elias 2007, Harris and Alcorn 2000-2001). Given that present procurement systems permit black markets to operate, how are these black markets changing? With a growing population in need of organs, greater ability to save lives through transplantation, and substantial unmet need, the black market is growing to fill the void.

World-systems theory divides countries into core, semi-peripheral, and peripheral countries. Core countries continue to benefit from a risk averse and laissez faire approach to organ governance (as contrasted to the riskier strategy involving regulation). The need is for transformative policy to intervene and promote an ethos of fairness and camaraderie that transcends national interest (Barnett, Beard, Kaserman 2002). If organ supplies can be increased legally, it follows that fewer people would resort to the black market.

In the case of a sales market the general concern is that peripheral countries will be crowded out of the market by wealthier core countries, simply because these actors have more money. What is to stop someone living in poverty in India from selling an organ to the highest bidder in the United States? How will people in India in need of organs compete with wealthier transplant recipients from other countries? Governments would need to cooperate to create a market with enough regulation that it provides fair and adequate compensation to organ suppliers while distributing organs to everyone in need, not just to the well to do.

The distinction between acquisition and allocation is noted here, because if the distribution of organs is regulated governments become the middlemen and gatekeepers by which goods are obtained or acquired early in the transference process, and allocated or apportioned later on in the process. Regulatory pressures could, as is the case in Iran, create a more equitable distribution of organs which does not depend on the ability to pay a premium. Where the policy is the best fit for the greatest number of individuals, the laws should follow.

## CONCLUSION

While a number of sociologists have studied organ procurement and recommended systems including the status quo, systems that make use of additional, appropriate compensation, and even systems that would legalize the outright sale of organs, there is little unanimity as to the direction in which we should head. Choosing a best policy is no simple task because it is difficult to predict future market behavior which is inherently uncertain; the accuracy of risk-assessment strategies is indeed fallible. Permitting the sale of organs perhaps has the potential to give life to those in need of organs and financial compensation to those willing to give. With so much at stake, so-called risk averse strategies are themselves risky.

However, a significant problem with predictive strategies which purport to solve the failures of any system is that the effects of a policy are inherently unpredictable and unknowable until that policy is implemented. Then, is the best policy the one which assumes the least uncertainty (e.g. the status quo)? The exact details about how a newly implemented organ procurement system would play out are beyond the scope and speculation of this paper. That said, the evidence that presumed consent systems work well (and would work well if implemented elsewhere) is especially noted because this approach provides a solution, in part, to the supply-side shortage without degrading the gift narrative. This is the approach I recommend – maintaining the gift narrative and organ donation system while adopting presumed consent to increase donations after death. I would also permit financial payments for donors' funeral expenses and their medical and other costs, including future costs, associated with donation and transplantation. Consider the negative implications of implementing a policy that permits the sale of organs. With the desacralization of the human body into distinct parts that can be sold, what is to prevent further degradation of the moral structure surrounding organ donations and a system which denies our basic notions of sacralization? The sale of organs would mark a paradigmatic shift after which the potential abuse and misuse of the laws could lead to even greater numbers using the black market, especially if the price of an organ is particularly high in the legal market.

Is the current organ procurement system a rational response to present conditions? On the one hand, the present system based on donations seems to operate within a morally acceptable framework. But its inability to come anywhere close to satisfying demand for transplant organs also gives rise to an independent black market, which in turn exploits poor people, enriches brokers and hospitals, and undermines the legal system. The most rational response should be one that results in the greatest amount of good for the greatest number of people. I encourage the addition of presumed consent and financial reimbursements as an effective solution to the ills of the present supply-side shortage. I do not recommend increasing financial incentives to such an extent that the organ donation is transformed from a gift to a commodity.

Institutions have influenced the economics of organ exchange, and through a process of trial and error, have developed a cultural account of altruistic donation designed to convince people to donate organs. These institutions have a continued and significant role in the organ procurement system. Altruism may have its limits, but the sale of organs would by no means guarantee improvement on such limits. I conclude that the sale of organs would likely fall short of the gift narrative, supported by the policy of presumed consent and appropriate financial incentives, in its ability and potential to limit both the use of the black market and exploitation of the poor while making progress to increase the much-needed supplies of legitimately obtained organs for all those in need.